\documentclass[preprint,12pt]{elsarticle}

\usepackage{amssymb}
\usepackage{subfigure}
\usepackage{booktabs}
\usepackage{threeparttable}
\usepackage{multirow}

\journal{XXX}

\begin{document}

\newcommand{\tabincell}[2]{\begin{tabular}{@{}#1@{}}#2\end{tabular}}

\begin{frontmatter}



\title{Autonomous quadrotor obstacle avoidance based on 
dueling double deep recurrent Q-learning with monocular vision}

\author[label1]{Jiajun Ou}
\author[label2]{Xiao Guo\corref{cor1}}
\ead{xiaoguo@buaa.edu.cn}
\author[label1]{Ming Zhu}
\author[label3]{Wenjie Lou}
\cortext[cor1]{Corresponding author}
\address[label1]{School of Aeronautic Science and Engineering, 
Beihang University, Beijing 100191, China}
\address[label2]{Research Institute for Frontier Science, 
Beihang University, Beijing 100191, China.}
\address[label3]{School of Electronic and Information Engineering, 
Beihang University, Beijing 100191, China}
    




\begin{abstract}
The rapid development of unmanned aerial vehicles (UAV) puts forward a
higher requirement for autonomous obstacle avoidance. Due to the limited
payload and power supply, small UAVs such as quadrotors usually carry simple
sensors and computation units, which makes traditional methods more
challenging to implement. In this paper, a novel framework is demonstrated
to control a quadrotor flying through crowded environments autonomously
with monocular vision. The framework adopts a two-stage architecture, consisting
of a sensing module and a decision module. The sensing module
is based on an unsupervised deep learning method. And the decision module
uses dueling double deep recurrent Q-learning to eliminate the adverse effects
of limited observation capacity of an on-board monocular camera. The
framework enables the quadrotor to realize autonomous obstacle avoidance
without any prior environment information or labeled datasets for training.
The trained model shows a high success rate in the simulation and a good
generalization ability for transformed scenarios.

\end{abstract}



\begin{keyword}
Unmanned aerial vehicle \sep
obstacle avoidance \sep
deep reinforcement learning

\end{keyword}

\end{frontmatter}


\section{Introduction}

Unmanned aerial vehicles(UAV) are widely used 
in both military and civil fields nowadays. 
UAVs can liberate people from monotonous or dangerous work scenarios 
such as searching and rescuing\cite{waharte2010supporting},  
package delivery\cite{song2018persistent} etc. 
However, considering the flight safety, 
the UAV operation depends on the human remote control
or follows a fixed flight route,
which may be labor-intensive and inefficient. 
With the increase of task complexity and scale,
it becomes imperative to develop autonomous flight ability.
To achieve autonomous flight,
the UAV needs to perceive the environment,
dealing with the environment information 
and avoiding obstacles in its expected flight path.
Typical on-board sensors for UAV are monocular camera, 
stereo camera, LIDAR, Kinect\cite{zhang2012microsoft} etc. 
While some kinds of sensors can output the depth information directly, 
such as LIDAR and Kinect, 
others get the depth information with additional calculations 
like a stereo camera.   

The quadrotor is widely used because of its low cost, 
flexibility and simple structure.
Theoretically, quadrotor can utilize sensor data 
and make proper action decisions to avoid obstacles. 
Nevertheless, with the restricted payload on it, 
the weight and energy consumption of equipped sensors are limited. 
In many cases, quadrotors can only afford 
to equip a fixed monocular camera,
providing limited environment observation only. 
Therefore, autonomous obstacle avoidance of quadrotor 
remains a challenging task.


Classical autonomous obstacle avoidance approaches
are based on Simultaneous Localization and Mapping(SLAM)\cite{murTRO2015, Engel2014LSD, montemerlo2003simultaneous}
or Structure from Motion (SfM)\cite{Wu2013Towards}.
Those approaches solve this problem with two separate technological processes, 
mapping and planning. 
Firstly they build a local map of surroundings based on sensor data, 
and then plan a path along with repetitively 
updating the local map\cite{lavalle1998rapidly, fox1997dynamic, ma2015local, ulrich1998vfh+}. 
SLAM and SfM based methods estimate the camera motion 
and depth by triangulation at each time step.
The critical step is the high-frequency feature extraction 
and matching in the reconstruction of the 3D local map 
from the sensor data.
Though the SLAM and SfM based approaches have been proven to be effective in autonomous obstacle avoidance, 
their disadvantages are apparent. 
The feature extraction may fail when facing an untextured obstacle 
and the real-time process requests unbearable computation 
for the on-board unit\cite{li2016real}.

The deep reinforcement learning method provides 
an alternative to autonomous obstacle avoidance\cite{silver2017mastering, levine2016end, mnih2015human}.
The methods based on deep reinforcement learning 
does not need feature extraction and matching at the pixel level, 
so it may be executed more efficiently.
However, their training requires sufficient groundtruth annotation data\cite{singla2019memory,gandhi2017learning},
which is expensive to obtain.
Moreover, preparing the training data with groundtruth
is even more effortful because 
the application scenario of quadrotors has considerable uncertainty.

In this paper, we propose a framework 
based on the novel deep reinforcement learning algorithm.
Our framework applies unsupervised learning based
depth estimation method to perceive surrounding obstacles.
It makes our framework no longer bothered 
by the preparation of annotated training data.
Since the on-board monocular camera of the quadrotor 
can only provide a limited field of view,
which may lead to the failure of traditional
reinforcement learning methods.
Our framework is built to make obstacle avoiding decisions 
according to the previous states 
rather than only the current one.
The trained model shows excellent performance in evaluation,
and its performance remains in the transformed scenarios.

The main contributions are as follows:

\begin{itemize}
    \item We propose a two-stage framework to 
    achieve quadrotor obstacle avoidance with monocular vision.
    A sensing module and a decision module are connected in series
    in our framework.
    The sensing module employs the unsupervised learning approach 
    to perform depth estimation,
    which takes view synthesis as the supervisory signal.
    So it can be trained by the raw image sequences captured by
    the on-board monocular camera.
    Thus, training the sensing module of our framework is free
    from the tedious preparing of data with groundtruth.
    \item We propose a dueling double 
    deep recurrent Q network to learn the obstacle avoidance policy.
    Since the image acquisition with fixed on-board monocular camera 
    can only provide limited field of view.
    It leads the quadrotor autonomous obstacles avoidance
    to become partial observable Markov decision process.
    Compared to other deep reinforcement models,
    our proposed network shows better learning efficiency 
    under partial observable conditions.
    \item We present a feasible solution for obstacle avoidance 
    in scenario transformation. 
    For the new scenarios from the same distribution, 
    the decision module of our framework shows a good generalization.
    Our decision module can maintain a high success rate 
    when dealing with new scenarios, 
    even though the appearance, size and location arrangement 
    of obstacles are different from the training scenario.

\end{itemize}

\section{Related work}

Learning-based avoidance methods can be 
divided into end-to-end architecture and hierarchical architecture.
The end-to-end architecture goes directly from sensor data
to obstacle avoidance actions.
Loquercio et al.\cite{loquercio2018dronet} design
a fast 8-layers residual network to output 
the steering angle and a collision
probability for each single input image.
The network is trained by the dataset
manually annotated by the authors.
Kouris et al.\cite{kouris2018learning} train 
convolutional neural networks(CNN) to predict 
distance-to-collision from the on-board monocular camera.
The proposed CNN is trained on the datasets annotated 
with real-distance labels, which are obtained by the 
Ultrasonic and Infra-Red distance sensors.
Moreover, Gandhi et al.\cite{gandhi2017learning} build a drone to sample 
data in the crash and their model 
learns a navigation policy from the sampled dataset.
To improve data efficiency, Zhu et al.\cite{zhu2017target} 
propose a novel simulation environment to train the model, 
which provides high-quality 3D scenes and a physics engine.
Although the end-to-end models can effectively avoid obstacles,
the training of these models needs a large number of data 
labeled with obstacle distance or collision probability, 
which requires much manual or special devices annotation.
Researchers take efforts to prepare these training data, 
which costs a lot of time and workforce.

On the other side, many researchers adopt
the hierarchical architecture to solve the monocular 
obstacle avoidance problem.
The typical hierarchical architecture contains two separate parts, 
environment sensing and decision making.
The monocular camera can only provide two-dimensional information directly,
and it is necessary to perceive three-dimensional 
information of the environment by utilizing 
depth estimation.
Supervised learning-based depth estimation 
achieved considerable 
results\cite{eigen2014depth, eigen2015predicting, long2015fully, laina2016deeper, hua2016depth}.
For solving the problem that the labeled datasets are difficult to obtain, 
researchers have proposed depth estimation methods 
based on semi-supervised learning\cite{kuznietsov2017semi} 
and unsupervised/self-supervised 
learning\cite{godard2017unsupervised, zhou2017unsupervised, yin2018geonet, chen2019self}.

Based on various monocular depth estimation methods, 
researchers have made progress in autonomous obstacle avoidance.
Tai et al.\cite{tai2016deep} build a highly compact network
structure which comprises
a CNN front-end network for perception and a fully connected
network for decision making.
The authors record the synchronized depth maps by Kinect and the
control commands by the human operator, and train the network
with supervised learning.
Sadeghi et al.\cite{sadeghi2016cad2rl} use the depth channel of Kinect 
to automatically annotate the RGB images with free-space/non-free-space labels,
proposing a learning method to train a fully convolutional neural network, 
which can be used to perform collision-free indoor flight in the real world.
In the paper\cite{xie2017towards}, a fully
convolutional neural network is constructed to predict depth from a raw RGB image, 
followed by a dueling architecture based deep double Q network for obstacle avoidance.
Singla et al.\cite{singla2019memory} use recurrent neural networks 
with temporal attention to realize UAV obstacle avoidance and autonomous exploration.
The authors train a conditional generative adversarial network 
to generate depth maps from RGB images.
For all these researches mentioned above,
their model training processes require data 
with labels or groundtruth.
In order to meet the needs of the training process for labeled data,
Yang et al.\cite{yang2019reactive} employ an online adaptive CNN 
for progressively improving depth estimation aided by monocular SLAM,
which increases the complexity of the system and the requirements of computation.
However, the application scenario of 
quadrotors is hard to restrict and predict in practice,
which increases the difficulty of data acquisition.
Therefore, considering the problem of data acquisition, 
the model training proposed in previous works 
is neither convenient nor economical, 
which may limit the practical application of these methods.

In order to reduce the difficulty of training 
and improve the feasibility of application, 
we present a novel framework 
to achieve autonomous obstacle avoidance in this paper.
The framework consists of two modules, and its training requires
no annotated datasets.
The first module is used for sensing the environment,
which adopts unsupervised learning based 
depth estimation to generate a depth map.
The second module responds to make obstacle avoidance decisions, 
whose policy is acquired through deep reinforcement learning. 
The former one can be trained by
raw RGB monocular image sequences,
and the latter one can be trained
in the simulation environment.
In this way, an autonomous obstacle avoidance method is proposed,
which is efficient and relatively easy to train.
In the face of unknown scenes, 
our framework only requires raw RGB image data to retrain, 
and then it can adapt to the new working scenario.

\section{Proposed method}

In this paper, a two-stage framework is proposed 
to sense the environment with an on-board monocular camera 
and make decisions to avoid obstacles in flight. 
This framework utilizes an unsupervised deep learning method
to estimate depth from the raw RGB monocular image. 
And the framework can further choose proper action 
to conduct safe flight without collision 
according to the generated depth information. 
The selected action acts on the outer loop control
of the quadrotor to realize the obstacle avoidance flight.
Our framework provides a feasible solution 
for obstacle avoidance with 
no prior environment information required.

\subsection{Problem definition}
The problem of autonomous obstacle avoidance for quadrotor 
can be reduced to Markov Decision Processes(MDPs), 
which can be defined as tuple 
$\langle \mathcal{S}, \mathcal{A}, \mathcal{T}\left(s_{t+1} | s_{t}, a_{t}\right), \mathcal{R}\left(s_{t}, a_{t}\right)\rangle$.
Here $\mathcal{S}$ is the set of states of the environment,
while $\mathcal{A}$ is the set of feasible actions.
$\mathcal{T}$ is the transition probability function and
$\mathcal{R}$ is the reward function.
At each time step $t$, the vehicle receives the state $s_t \in \mathcal{S}$
and propose action $a_t \in \mathcal{A}$.
And the received reward $r_t$ is given by the reward function $\mathcal{R}\left(s_{t}, a_{t}\right)$. 
In accordance with the transition model $\mathcal{T}\left(s_{t+1} | s_{t}, a_{t}\right)$, 
the vehicle moves into a new state $s_{t+1}$. 
The action $a_t$ is sampled from the policy $\pi = P(a_t|s_t)$. 
The expectation of accumulative reward can 
be approximated by action-state-value function $Q(s_t,a_t)$, 
which is constructed by a deep neuron network.

The key to this problem is to find the optimal policy $\pi$ to 
maximize the the accumulative 
future reward 
$\mathbb{E}\left[\sum_{t}^{\infty} \gamma^{t} R\left(s_{t}, a_{t}\right)\right]$, 
where $\gamma$ is the discount factor.
By choosing the optimal action which maximizes the Q-value each time, 
the optimal Q-value function can be computed using the Bellman equation
\begin{equation}
    Q^{*}\left(s_{t}, a_{t}\right)=\mathbb{E}_{s_{t+1}}\left[r+\gamma \max _{a_{t+1}} Q^{*}\left(s_{t+1}, a_{t+1}\right) | s_{t}, a_{t}\right]
\end{equation}
The optimal policy is capable of leading the quadrotor 
to make correct action decision to avoid obstacles during flight.

In this paper, we simplify the obstacle avoidance problem by fixing the flight altitude and forward speed. 
And the framework only controls the flight direction by adjusting the yaw angular rate of the quadrotor. 
The framework uses the image obtained by the monocular camera, which fixed on the quadrotor 
at each time step to output the proper flight direction, realizing autonomous navigation without prior obstacle information.

\subsection{Sensing with unsupervised depth estimation}

The sensing module of our framework constructs a fully connected 
neural network, which is capable of 
mapping directly from the input RGB images 
to the estimate of the underlying scene structure.
It employs the DispNet\cite{mayer2016large} 
to generate a front view depth map. 
Inspired by the work in\cite{zhou2017unsupervised},
the network is trained by  the supervision signal 
that the task of novel view synthesis generates. 
The training  process only requires the raw RGB image sequences 
obtained by the on-board monocular camera while the quadrotor is flying.

The captured image sequences are stored in the replay buffer.
The target image and two nearby images in the sequences are sampled 
from the replay buffer randomly.
These images are input to the depth estimation network 
and pose network at the same time.
The depth estimation network generates the depth map $\widehat{D}_t$ from the target image. 
The pose network takes both the target image $I_t$ 
and the nearby images $I_s$ ($I_{t-1}$ and $I_{t+1}$) in the sequence as input, 
and outputs the relative camera poses $\widehat{T}_{t\rightarrow s}$
($\widehat{T}_{t\rightarrow t-1}$ and $\widehat{T}_{t\rightarrow t+1}$). 
The photometric reconstruction loss between the raw target image 
and the reconstructed target image is used for training the networks,
which can be defined as follows
\begin{equation}
    \mathcal{L}_{vs}=\sum_{s} \sum_{p}\left|I_{t}(p)-\hat{I}_{s}(p)\right|,
\end{equation}
where $p$ represents the index over pixel coordinates,
the $I_t$ is the raw target image, 
and $\hat{I}_{s}$ is the synthesis target image 
warped from nearby image.

To reconstruct $I_t$, pixels are sampled
from $I_s$ based on the depth map $\widehat{D}_t$
and the relative pose $\widehat{T}_{t\rightarrow s}$.
$p_t$ is projected coordinates onto $p_s$ can be obtained
as follows
\begin{equation}
    p_{s} \sim K \hat{T}_{t \rightarrow s} \hat{D}_{t}\left(p_{t}\right) K^{-1} p_{t},
\end{equation}
where $p_t$ represents the coordinates of a pixel
in the target view, and K represents the camera intrinsics
matrix.

\begin{figure}[htbp]
    \centering
    \includegraphics[width=12cm]{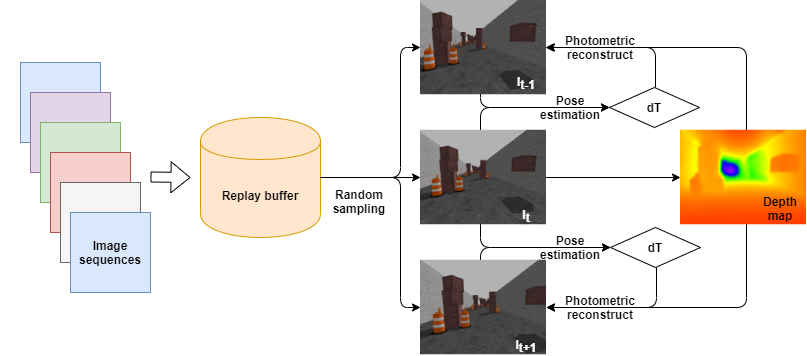}
    \caption{Unsupervised learning based on view synthesis}
    \label{fig:dispnet}
\end{figure}

By utilizing view synthesis as supervision, 
the depth estimation network is trained in an unsupervised manner 
from captured image sequences.
The training pipeline is shown in Figure \ref{fig:dispnet}.

\subsection{Dueling double deep recurrent Q network}
\begin{figure}[htbp]
    \centering
    \subfigure[Top-down view in the simulation]{
        \begin{minipage}{6cm}
        \centering
        \includegraphics[width=6cm,height=4cm]{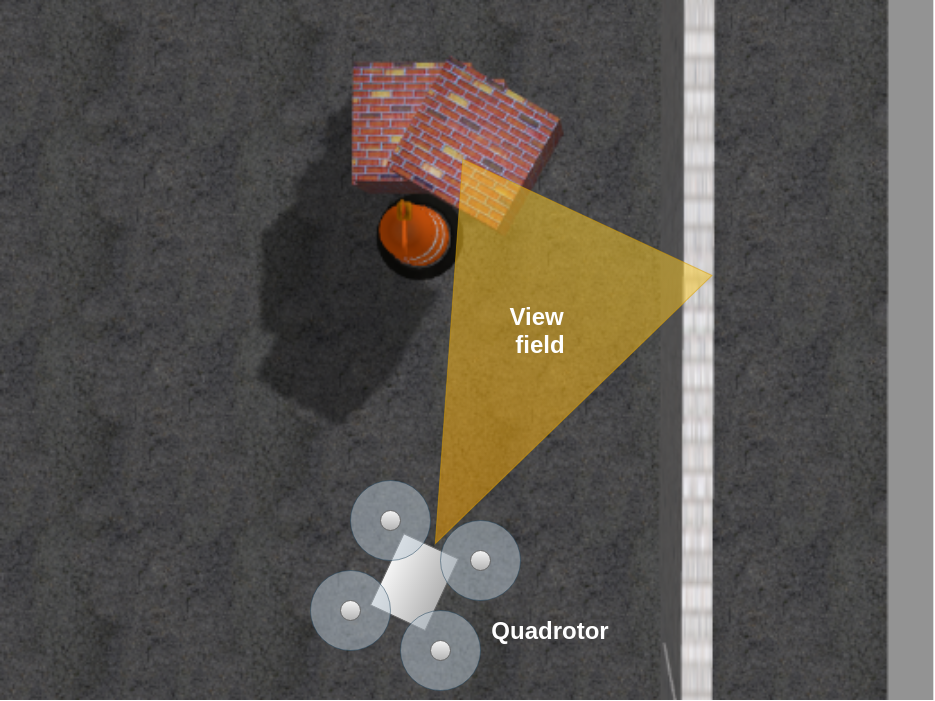}
        \end{minipage}
    }
    \subfigure[Front view depth map observed]{
        \begin{minipage}{6cm}
        \centering
        \includegraphics[width=6cm,height=4cm]{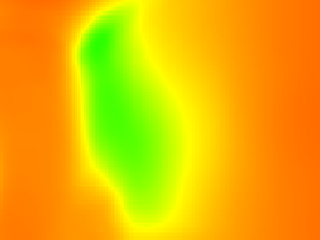}
        \end{minipage}
    }
    \caption{A demonstration of collision cased by partial observation}
    \label{fig:corner}
\end{figure}

The on-board monocular camera can only 
provide a limited field of view of the surrounding environment.
The partial observability makes it hard to gain the optimal policy 
in some particular scenes\cite{singla2019memory}.
As it is shown in Figure \ref{fig:corner},
the quadrotor might fly straight forward 
and crash on the obstacle based on 
the current partial observation,
while the proper action is turning left.

\begin{figure}[htbp]
    \centering
    \subfigure[Raw image]{
        \begin{minipage}{6cm}
        \centering
        \includegraphics[width=6cm,height=4cm]{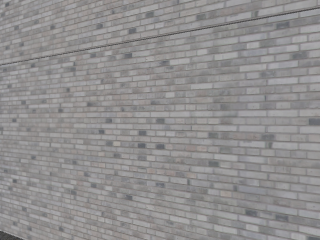}
        \end{minipage}
    }
    \subfigure[Depth map]{
        \begin{minipage}{6cm}
        \centering
        \includegraphics[width=6cm,height=4cm]{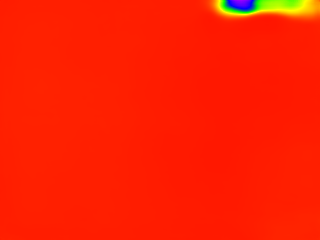}
        \end{minipage}
    }
    \caption{Poor performance of depth estimation in some scenes}
    \label{fig:bad case}
\end{figure}

Besides, the training data of the depth estimation network is 
captured by the on-board camera
of the quadrotor while it is flying.
Unsupervised depth estimation method has been proven feasible, 
but using on-board camera data to train the model may
raise a new problem.
The limitation of the quadrotor's flight ability and the avoidance 
of crash lead the data to be short of comprehensiveness.
And it may cause poor depth estimation performance in some scenes,
shown in Figure \ref{fig:bad case}.

Considering the above situations,we treat the 
quadrotor obstacle avoidance
as partially observable Markov decision processes(POMDPs)
in this paper.
The POMDP problem can be defined as tuple
$\langle\mathcal{S}, \mathcal{A}, \mathcal{T}, \mathcal{R}, \Omega, \mathcal{O}\rangle$.
$\mathcal{S}, \mathcal{A}, \mathcal{T}, \mathcal{R}$ are respectively
the set of states, actions, transitions, and rewards as before.
Here $\Omega$ is the set of observations, 
while $\mathcal{O}$ is the set of the probability distributions.
The on-board monocular camera gets observations $o \in \Omega$
generated from the underlying system state according
to the probability distribution $o \sim \mathcal{O}(s)$.
At current time $t$, the observation $o_t$ can only represent part of 
the current surrounding environment state $s_t$.
Since estimating Q-value $Q(o_t, a_t | \pi) \neq Q(s_t, a_t | \pi)$, 
obstacle avoidance action $a_t$ relying entirely on 
current observation may be fragile.
Therefore, in this paper, a method that can use 
previous observation experience is proposed
to make decisions for improving the performance of obstacle avoidance.
The model is able to extra useful environment information
from sequential observation before the current time,
making $Q(o_t, a_t | \pi)$ closer to $Q(s_t, a_t | \pi)$.
So it can eliminate the interference of low-quality observation results,
and avoid quadrotors getting trapped in cases like the example
in Figure \ref{fig:corner}.

\begin{figure}[htbp]
    \centering
    \includegraphics[width=12cm]{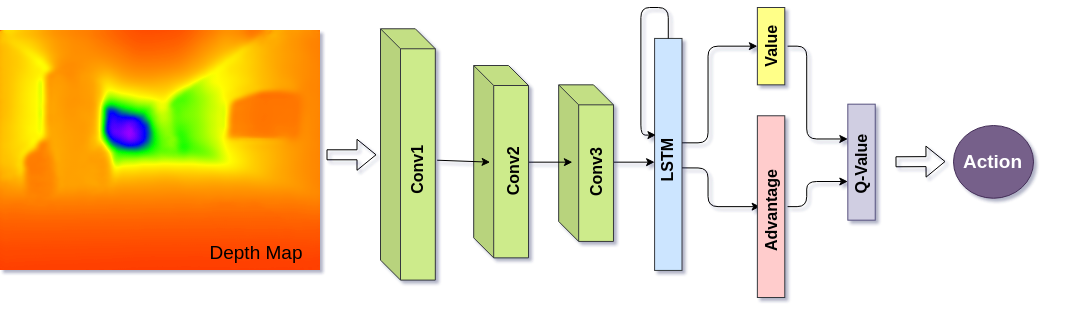}
    \caption{The structure of dueling deep recurrent Q network}
    \label{fig:DDRQN}
\end{figure}

The model is based on the deep recurrent Q network\cite{hausknecht2015deep}(DRQN) 
with the dueling and double technology\cite{wang2015dueling, van2016deep}.
In the traditional dueling network, two streams are used
to compute the value and advantage functions. 
The dueling network can improve performance and training speed.
On the other hand, the double technology solving the problem of
overoptimistic value estimation.
Based on these previous research results, 
we combine the DRQN and dueling network by
replacing one fully connected layer 
of the dueling network with an LSTM\cite{hochreiter1997long} layer.
This dueling deep recurrent Q network structure
is shown in Figure \ref{fig:DDRQN}, 
and corresponding parameters are shown in Table \ref{tab:1}.

\begin{table}[htbp]

    \caption{Parameters of the dueling deep recurrent Q network}    
    \centering
    \begin{threeparttable} 
    \label{tab:1}      
    \begin{tabular}{c|ccc}      
    \hline\noalign{\smallskip}
    Item & \tabincell{c}{Size\\(height,width,channel)} & Number & Stride  \\
    \noalign{\smallskip}\hline\noalign{\smallskip}
    Depth map & (128,416,1) & - & - \\
    Conv 1 & (8,8,4) & - & 4 \\
    Conv 2 & (4,4,8) & - & 2 \\
    Conv 3 & (3,3,8) & - & 2 \\
    LSTM & - & 1152 & - \\
    FC for advantage & - & 5 & - \\
    FC for value & - & 1 & - \\
    \noalign{\smallskip}\hline
    \end{tabular}
    \end{threeparttable}  
\end{table}

\section{Training and Testing}
\begin{figure}[htbp]
    \centering
    \subfigure{
    \begin{minipage}{6cm}
    \centering
    \includegraphics[width=6cm,height=4cm]{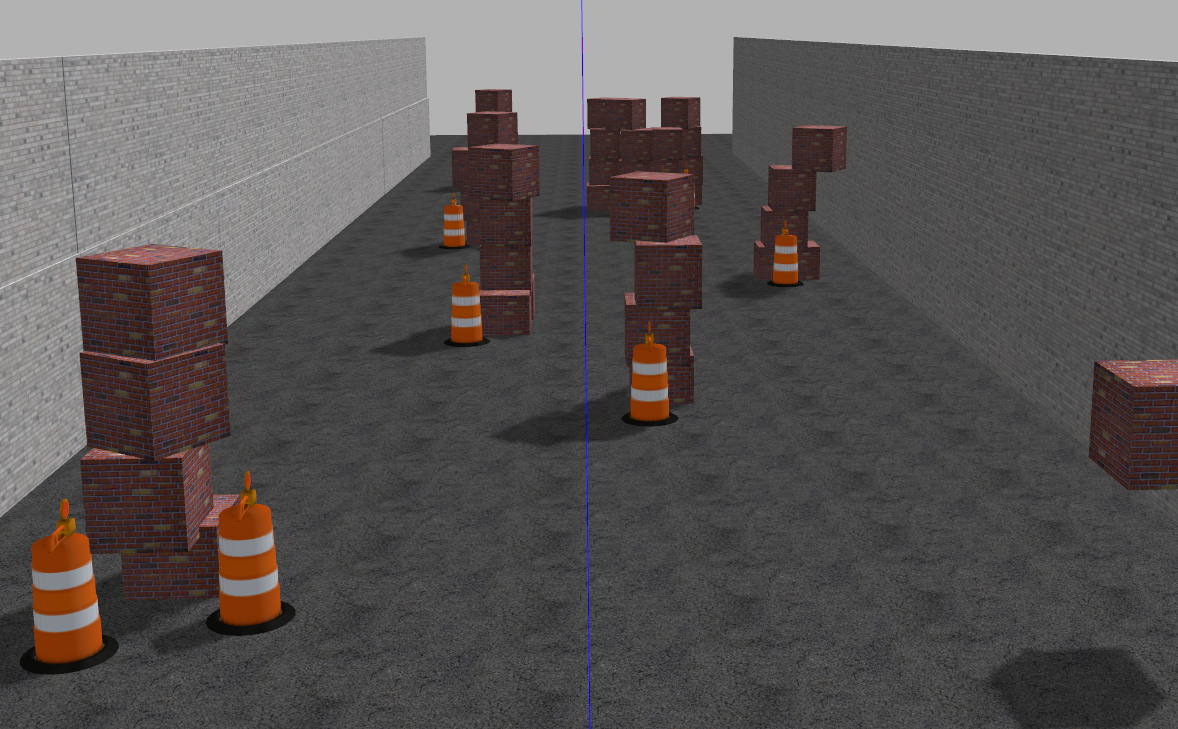}
    \end{minipage}
    }
    \subfigure{
    \begin{minipage}{6cm}
    \centering
    \includegraphics[width=6cm,height=4cm]{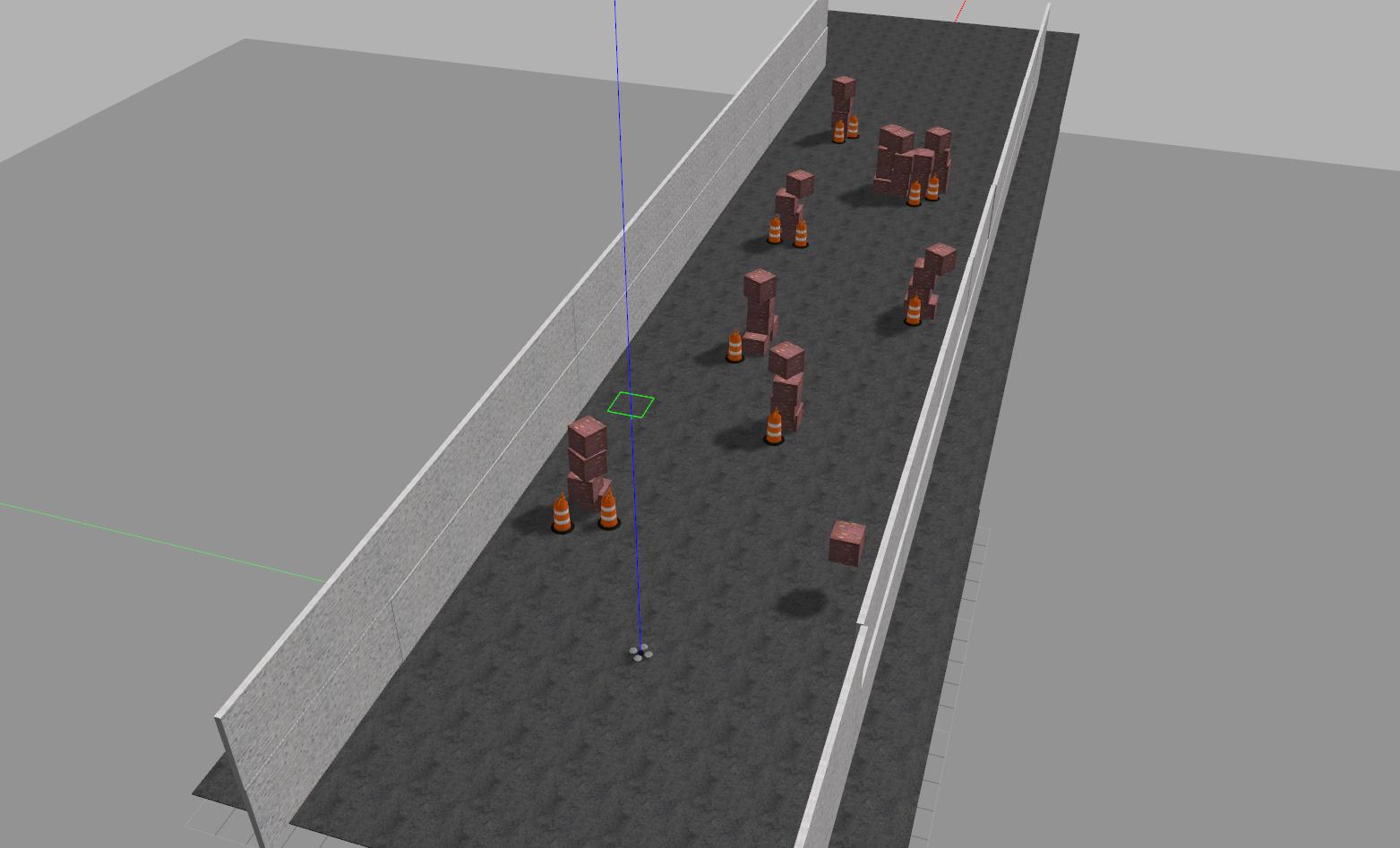}
    \end{minipage}
    }
\caption{The basic training environment in Gazebo}
\label{fig:simulation env}
\end{figure}

The model is trained in the Gazebo simulation environment 
with a step-by-step training strategy.
The depth estimation network is firstly trained.
Then the trained depth estimation network and the depth maps
it generates are used to train 
the dueling double deep recurrent Q network(D3RQN),
wich is responsible for performing
obstacle avoidance decision making.
Several models are trained and evaluated in multiple different
simulation environments. 
Figure \ref{fig:simulation env} shows the screenshots of
the basic training environment in simulation.


\subsection{Depth estimation network}

The image collection process is conducted 
in the simulation environment by the on-board monocular camera
of the manually controlled quadrotor.
These images are used to train the depth estimation network.
The training hyper-parameters
are shown in Table \ref{tab:2}.
\begin{table}[htbp]
    \caption{Parameters of training the depth estimation network}    
    \centering
    \label{tab:2}      
    \begin{tabular}{cc}      
    \hline\noalign{\smallskip}
    Parameters & Value \\
    \noalign{\smallskip}\hline\noalign{\smallskip}
    Image number & 5000\\
    Batch size & 4 \\
    Learning rate & 0.00005 \\
    Image acquisition interval & 0.4s\\
    Camera linear velocity & 2m/s \\
    Training iteration & 30000\\

    \noalign{\smallskip}\hline
    \end{tabular}
\end{table}

\begin{figure}[htbp]
    \centering
    \subfigure[Raw images]{
        \begin{minipage}{4cm}
        \centering
        \includegraphics[width=4cm,height=2.67cm]{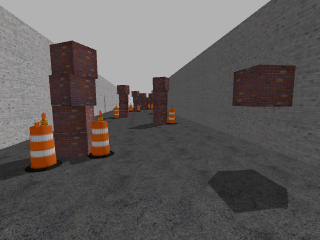}
        \end{minipage}
        \begin{minipage}{4cm}
        \centering
        \includegraphics[width=4cm,height=2.67cm]{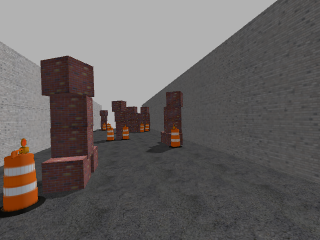}
        \end{minipage}
        \begin{minipage}{4cm}
        \centering
        \includegraphics[width=4cm,height=2.67cm]{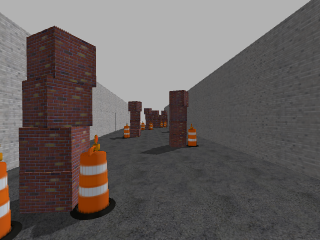}
        \end{minipage}
    }
    \subfigure[Depth maps]{
        \begin{minipage}{4cm}
        \centering
        \includegraphics[width=4cm,height=2.67cm]{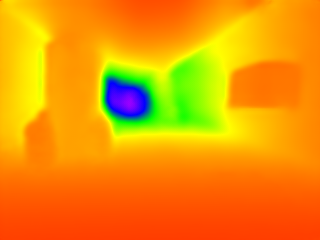}
        \end{minipage}
        \begin{minipage}{4cm}
        \centering
        \includegraphics[width=4cm,height=2.67cm]{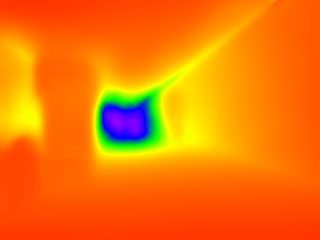}
        \end{minipage}
        \begin{minipage}{4cm}
        \centering
        \includegraphics[width=4cm,height=2.67cm]{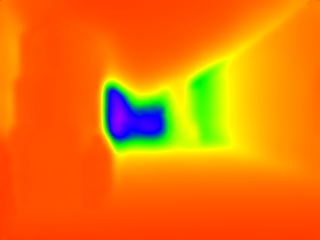}
        \end{minipage}
    }
    \caption{Examples of depth estimation performance (Red: near, Blue: far)}
    \label{fig:depth prediction demo}
\end{figure}

The depth estimation network is evaluated after training 30000 iterations,
which is much less than that in the original paper\cite{zhou2017unsupervised}.
The examples of depth estimation are shown in Figure \ref{fig:depth prediction demo}.
And we test the depth estimation network on an NVIDIA GeForce RTX 2070 GPU with 8 GB RAM
and Intel Core i7 processor machine, and the depth map generation rate
reaches more than 30Hz.

\subsection{Dueling double deep recurrent Q network}
\begin{figure}[htbp]
    \centering
    \subfigure[DDQN]{
        \begin{minipage}{6cm}
        \centering
        \includegraphics[width=6cm,height=4cm]{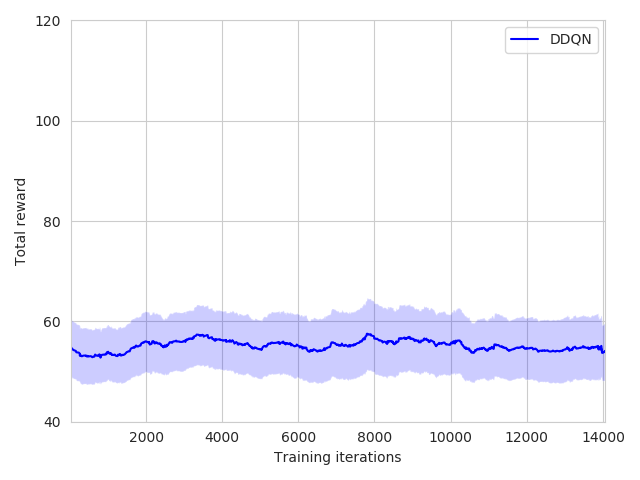}
        \end{minipage}
    }
    \subfigure[D3QN]{
        \begin{minipage}{6cm}
        \centering
        \includegraphics[width=6cm,height=4cm]{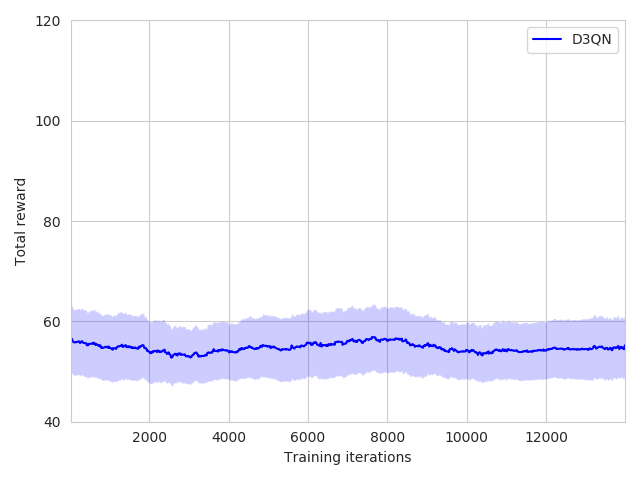}
        \end{minipage}
    }
    \subfigure[DDRQN]{
        \begin{minipage}{6cm}
        \centering
        \includegraphics[width=6cm,height=4cm]{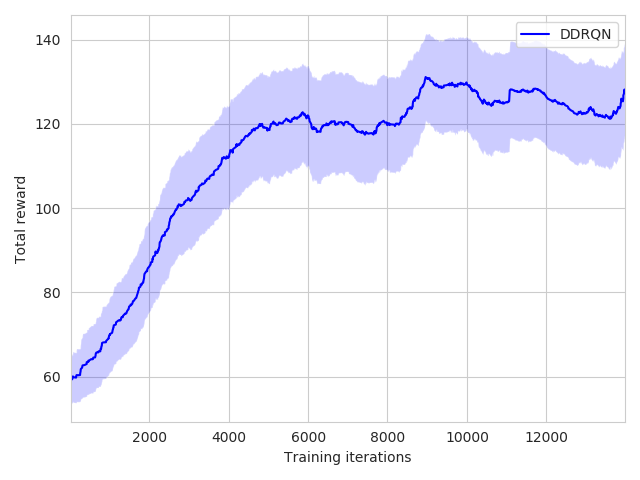}
        \end{minipage}
    }
    \subfigure[D3RQN]{
        \begin{minipage}{6cm}
        \centering
        \includegraphics[width=6cm,height=4cm]{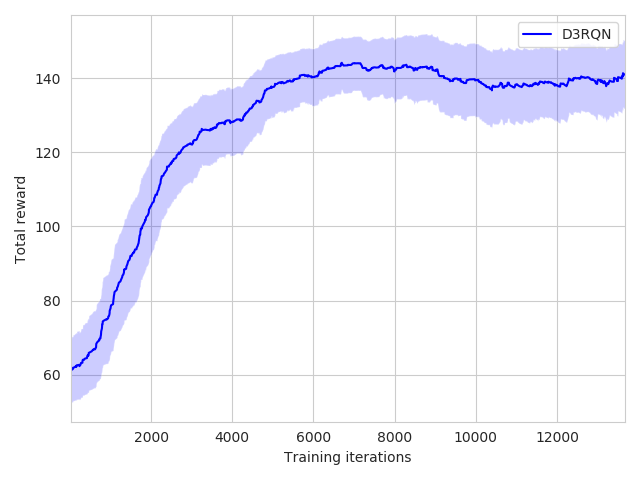}
        \end{minipage}
    }
    \caption{Learning curves of the DDQN, D3QN, DDRQN and our D3RQN.}
    \label{fig:learning curves}
\end{figure}

\begin{figure}[htbp]
    \centering
    \includegraphics[width=12cm]{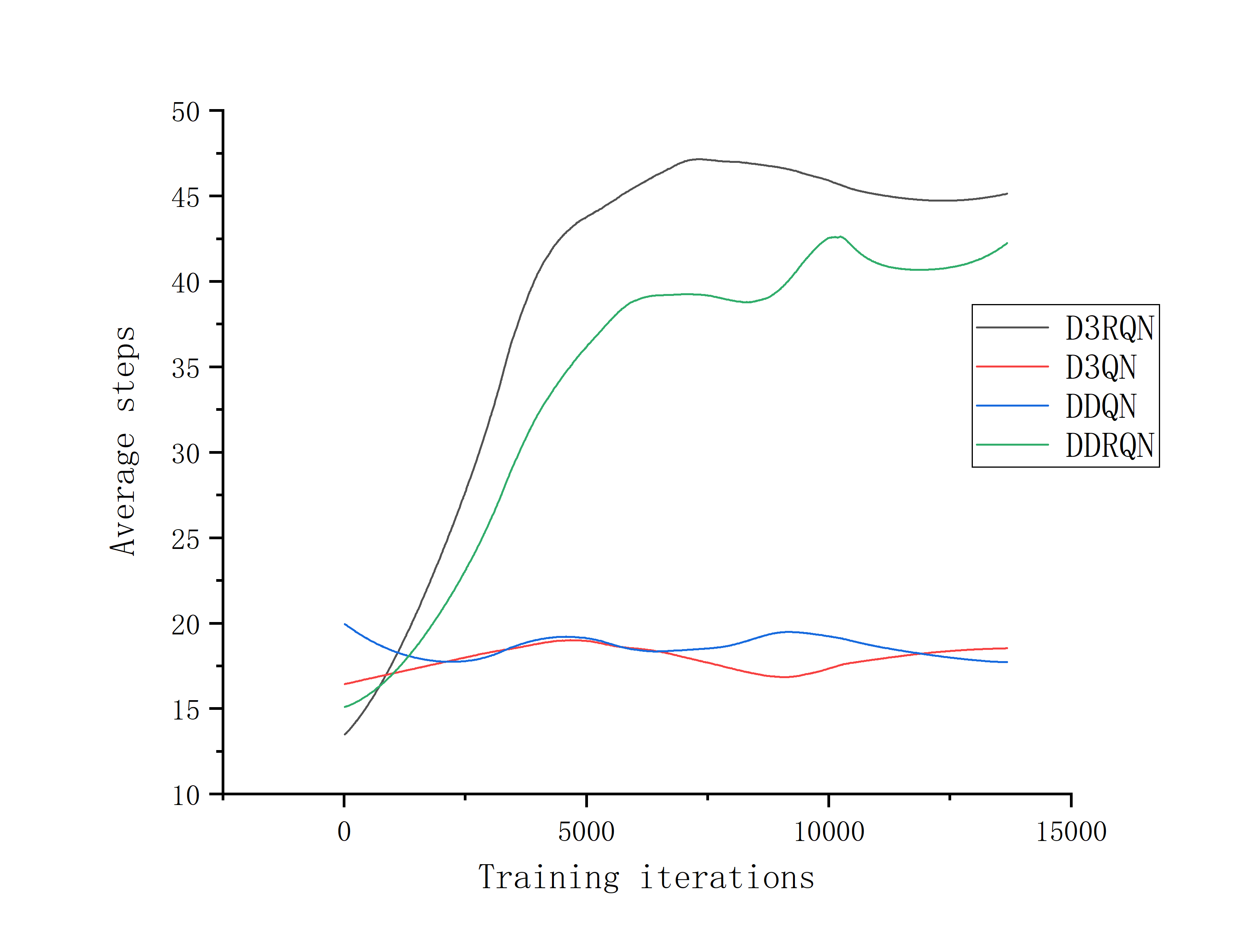}
    \caption{The comparison of the four methods after smoothing}
    \label{fig:compare}
\end{figure}

The dueling double deep recurrent Q network 
is also trained in the simulation environment for cost and safety reasons.
The network is trained to estimate 
current Q-value 
over the last several observations,
which means the last several depth maps
generated by the depth estimation network.
The obstacle avoidance includes five actions, 
which are defined in Table \ref{tab:3}.
With each action, the quadrotor obtains
a reward which defined as

\begin{equation}
    R=\left\{\begin{array}{lr}
    d_{nearest} & d_{nearest}\geq 0.5 \\
    -1 & d_{nearest}<0.5
    \end{array}\right.
\end{equation}
where the $d_{nearest}$ is the distance to the nearest obstacle and the safe distance is 0.5. When the $d_{nearest}$ is smaller than the safe distance,
the collision is considered to happen and 
the training episode ends obtaining a negative reward.
Besides our dueling double deep recurrent Q network, 
three other RL based models are trained in the simulation environment 
along with similar parameters.
They are double deep Q network(DDQN), dueling double deep Q network(D3QN) 
and double deep recurrent Q network(DDRQN).
The learning curves of the four models are shown in Figure \ref{fig:learning curves}.
Figure \ref{fig:compare} presents the comparison of different models. 
All the data preparing and training processes are 
on an NVIDIA GeForce RTX2070 machine.

\begin{table}[htbp]
    \caption{Action definition of the quadrotor}    
    \centering
    \label{tab:3}      
    \begin{tabular}{c|cc}      
    \hline\noalign{\smallskip}
    Action num & \tabincell{c}{linear velocity \\(x,y,z)}  &\tabincell{c}{angular velocity \\(x,y,z)} \\
    \noalign{\smallskip}\hline\noalign{\smallskip}
    1 & (2,0,0)&(0,0,0) \\
    2 & (2,0,0)&(0,0,0.25) \\
    3 & (2,0,0)&(0,0,0.25) \\
    4 & (2,0,0)&(0,0,0.5) \\
    5 & (2,0,0)&(0,0,0.5) \\

    \noalign{\smallskip}\hline
    \end{tabular}
\end{table}

\begin{table}[htbp]
    \caption{Parameters of training the dueling double deep recurrent Q network}    
    \centering
    \label{tab:4}      
    \begin{tabular}{cc}      
    \hline\noalign{\smallskip}
    Parameters & Value \\
    \noalign{\smallskip}\hline\noalign{\smallskip}
    Batch size & 32 \\
    Discount factor& 0.99 \\
    Learning rate & 0.0003 \\
    Input sequence length & 5 \\
    Action time interval & 0.4s\\
    Target network update frequency & 300 \\

    \noalign{\smallskip}\hline
    \end{tabular}
\end{table}

The four RL based models are tested in the Gazebo environment. 
In the test, once the model controls the quadrotor 
to fly more than 50 steps without collision,
we consider it is a success.
These models are evaluated by calculating the success rate of each model
in 2000 times test flight, the results are shown in Table \ref{tab:5}.
And in the test, our whole framework can run on the machine mentioned before at more than 15 Hz.
\begin{table}[htbp]
    \caption{Test results of 4 different models}    
    \centering
    \label{tab:5}      
    \begin{tabular}{ll}      
    \hline\noalign{\smallskip}
    Model  & Success rate  \\
    \noalign{\smallskip}\hline\noalign{\smallskip}
    Straight     & 0     \\
    Random       & 0.002 \\
    DDQN         & 0.137 \\
    D3QN         & 0.152 \\
    DDRQN        & 0.673 \\
    Our approach & 0.994 \\ 
    \noalign{\smallskip}\hline
    \end{tabular}
\end{table}

\subsection{Performance after scenario transformation}

Since the scenario uncertainty in UAV applications
is usually strong,
the quadrotor obstacle avoidance ability
should be effective in different scenarios.
Previous researches have focused on building 
complex models to adapt to different scenarios as much as possible.
However, it is hard for training datasets to cover 
all possible scenario types.
And a complex model is not suitable for 
the airborne processing unit of a quadrotor.
Rather than solving all problems in one model,
our approach is dedicated to realizing 
more convenient training
when facing new application scenarios.

\begin{figure}[htbp]
    \centering
    \subfigure[Raw image]{
        \begin{minipage}{6cm}
        \centering
        \includegraphics[width=6cm,height=4cm]{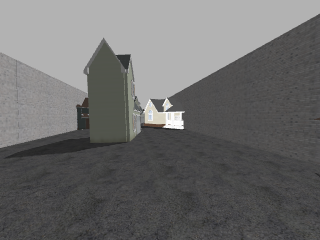}
        \end{minipage}
    }
    \subfigure[Depth map]{
        \begin{minipage}{6cm}
        \centering
        \includegraphics[width=6cm,height=4cm]{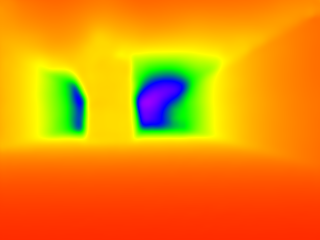}
        \end{minipage}
    }
    \caption{Depth prediction performance after scenario transformation}
    \label{fig:tansfer_depth}
\end{figure}

In this section, new simulation environments 
are used to test our model.
The appearance, size, shape and location of obstacles 
in the new simulation environments 
are different from those 
in the basic environment in Figure \ref{fig:simulation env}.
The only thing needed to do before conducting the test is 
retraining the depth estimation network with image sequences
obtained in the new environments.
The depth estimation performance after scenario transformation
is shown in Figure \ref{fig:tansfer_depth}.

\begin{figure}[htbp]
    \centering
    \subfigure[Env-1: the narrow channel]{
        \begin{minipage}{6cm}
        \centering
        \includegraphics[width=6cm,height=4cm]{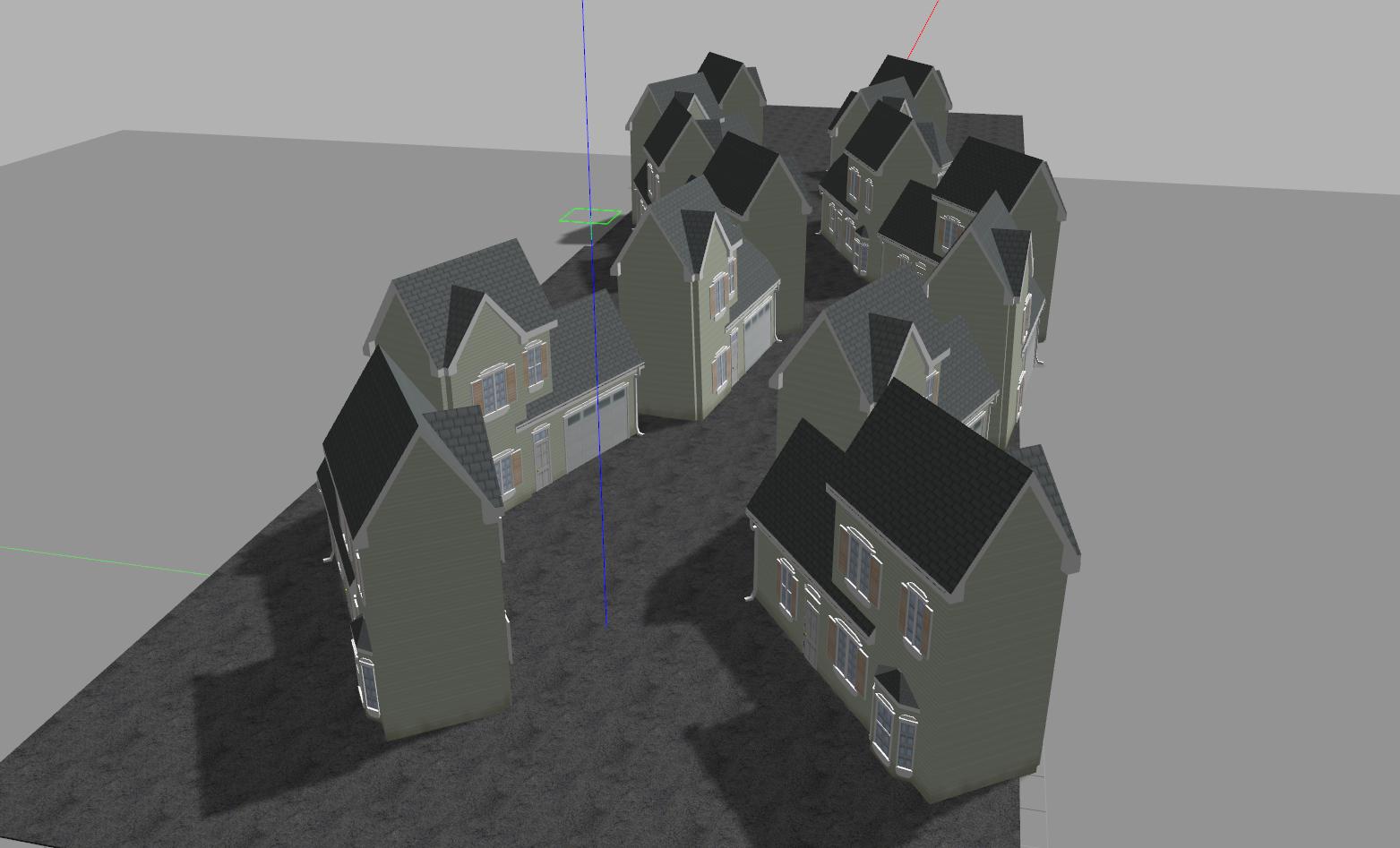}
        \end{minipage}
        \begin{minipage}{6cm}
        \centering
        \includegraphics[width=6cm,height=4cm]{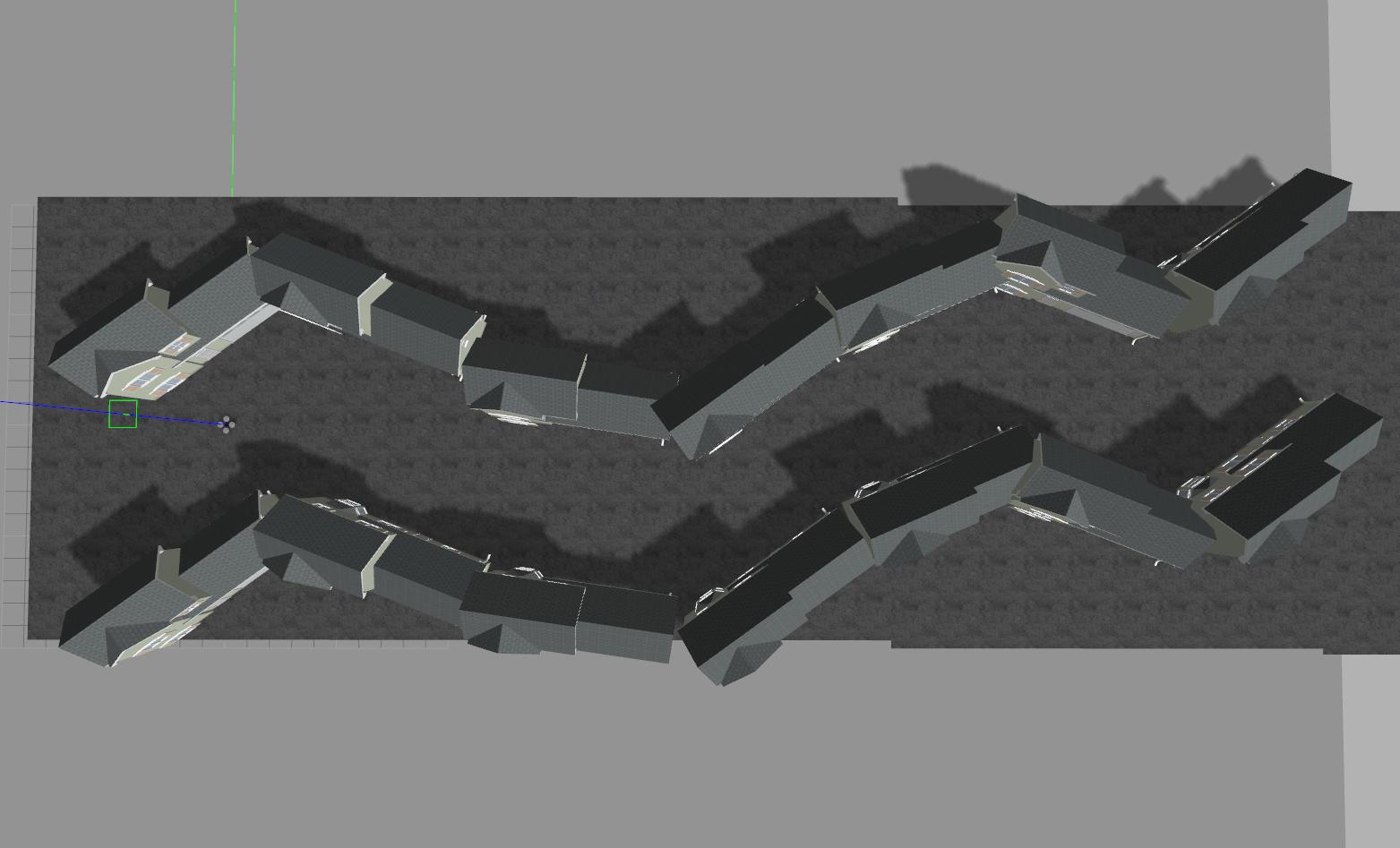}
        \end{minipage}
    }

    \subfigure[Env-2: the intersections]{
        \begin{minipage}{6cm}
        \centering
        \includegraphics[width=6cm,height=4cm]{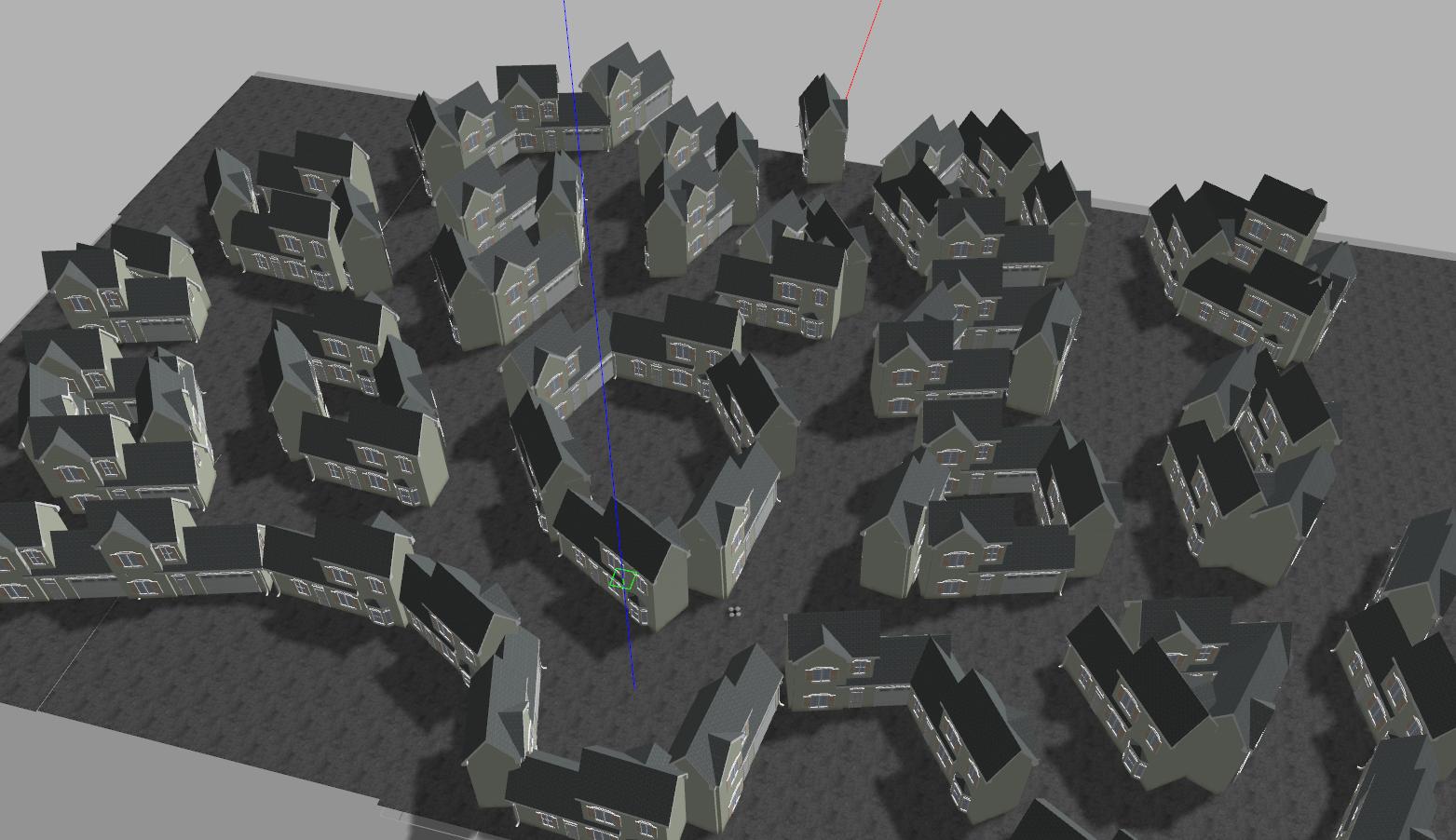}
        \end{minipage}
        \begin{minipage}{6cm}
        \centering
        \includegraphics[width=6cm,height=4cm]{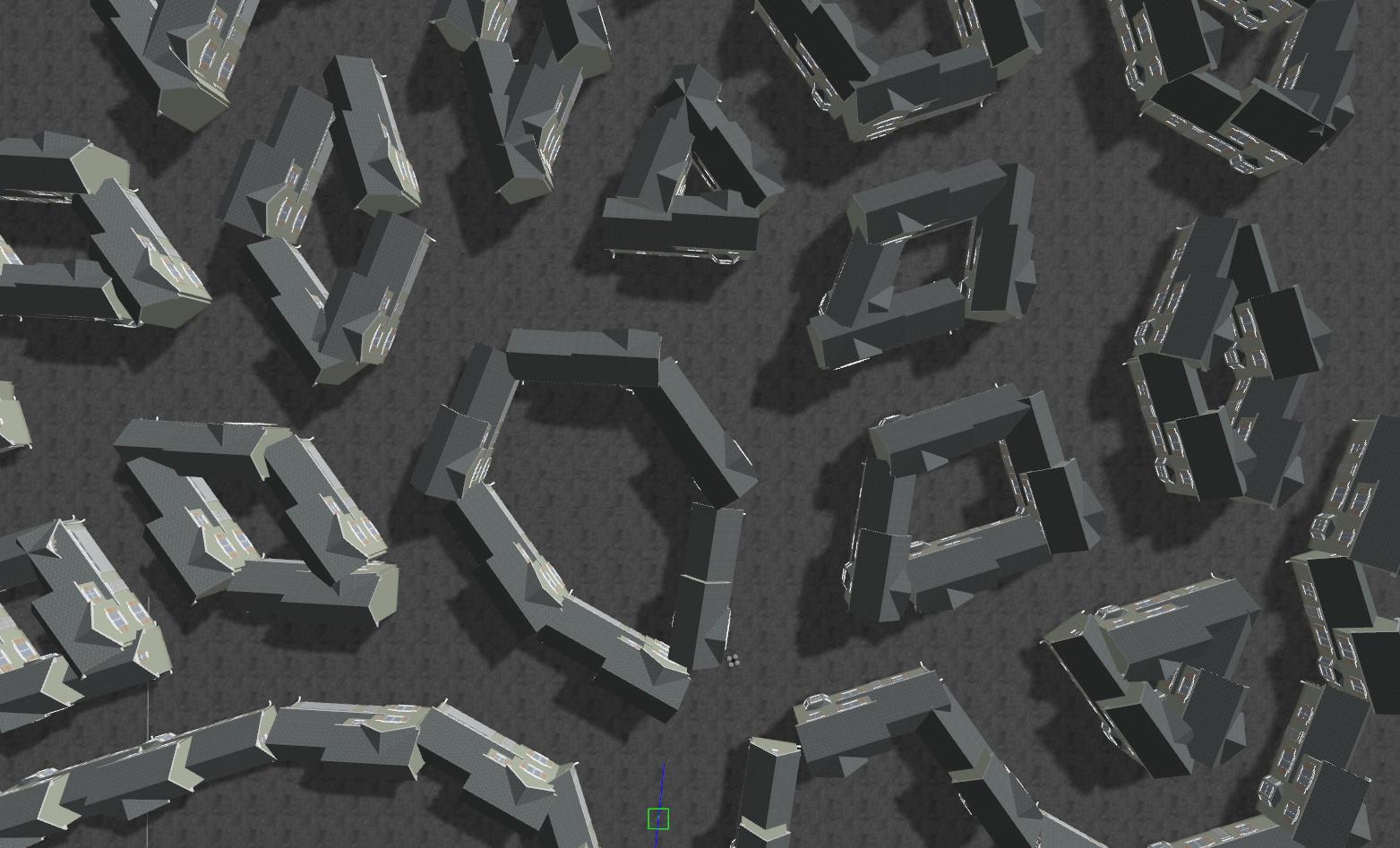}
        \end{minipage}
    }

    \subfigure[Env-3: the corners]{
        \begin{minipage}{6cm}
        \centering
        \includegraphics[width=6cm,height=4cm]{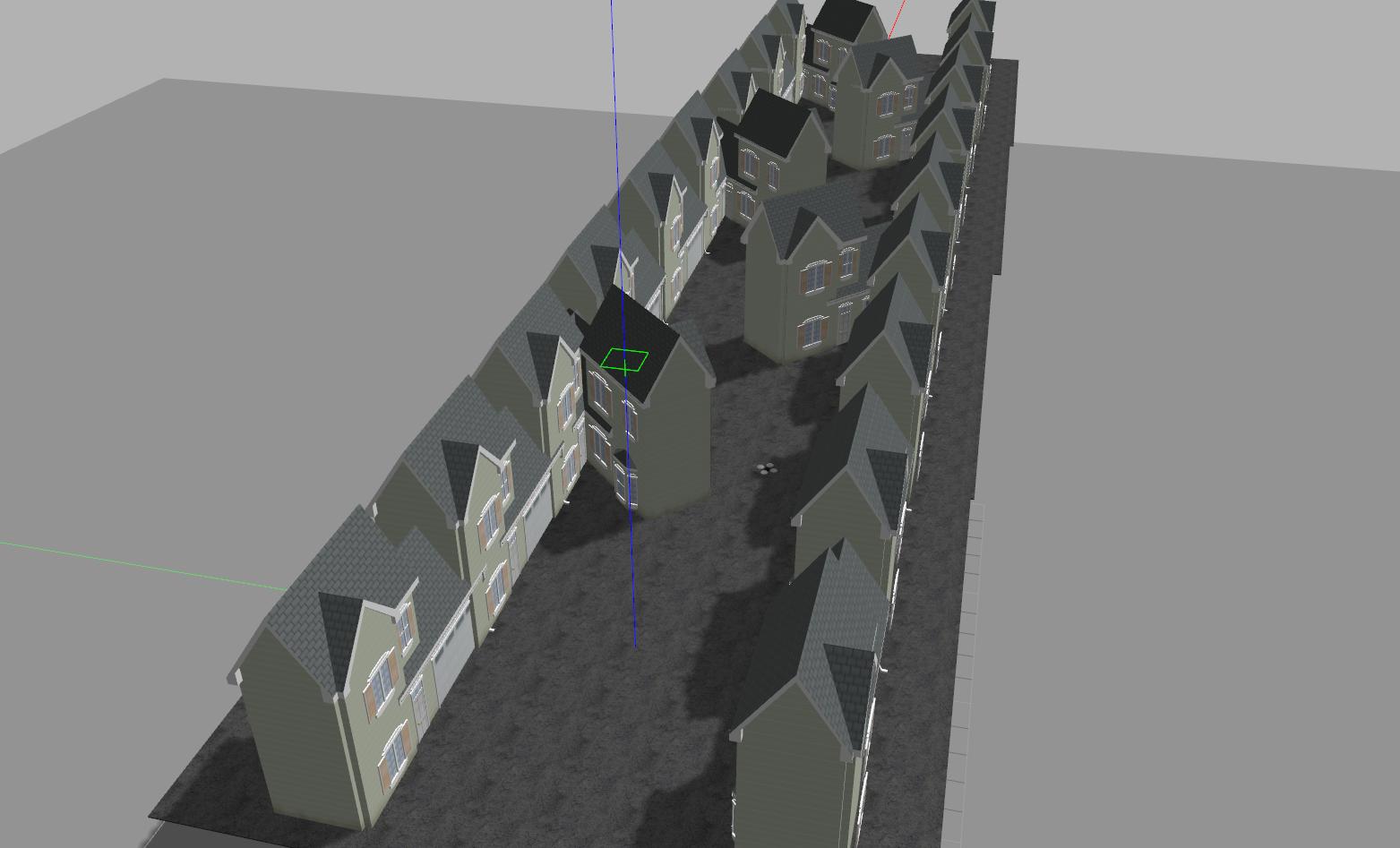}
        \end{minipage}
        \begin{minipage}{6cm}
        \centering
        \includegraphics[width=6cm,height=4cm]{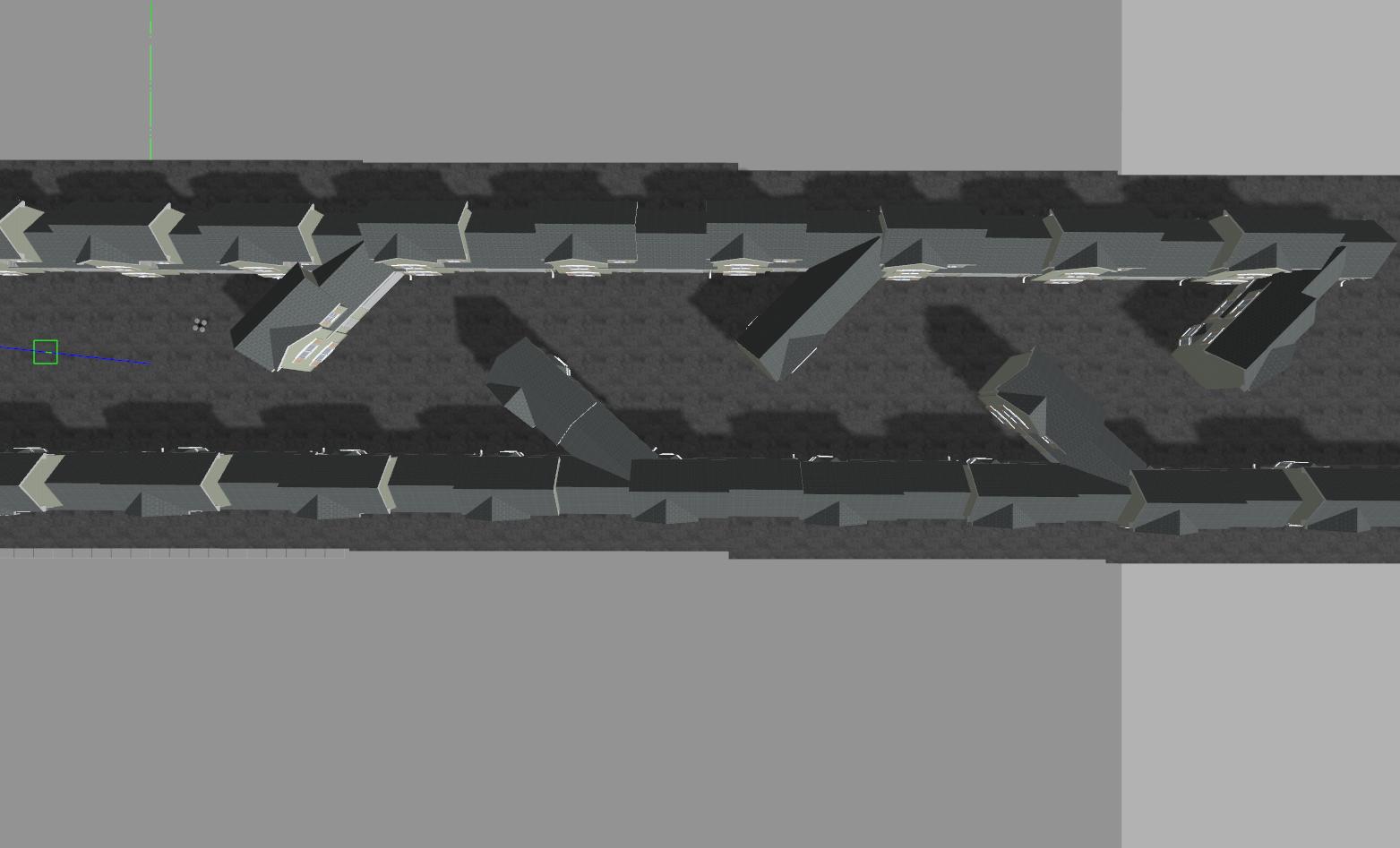}
        \end{minipage}
    }
    \caption{The screenshots of the test enviroments}
    \label{fig:tansfer_simu_env}
\end{figure}

With the retrained depth estimation network, 
the whole model is tested in new simulation environments.
The screenshots of the test environments 
are shown in Figure \ref{fig:tansfer_simu_env}.
It's worth emphasizing that we reuse 
the dueling double deep recurrent Q network trained 
in the basic environments
without any fine-tune operation.
These simulation scenarios respectively represent narrow channels, 
intersections and corners.
And Table \ref{tab:performance_transfer} presents 
the performance of our method after the scenario transformation.

\begin{table}[htbp]  
    \centering  
    \caption{Test results of obstacle avoidance after scenario transformation} 
    \begin{threeparttable}  
    \label{tab:performance_transfer}  
      \begin{tabular}{cccc}  
      \toprule  
      \multirow{2}{*}{Model}&  \multicolumn{3}{c}{Success rate}\cr  
      \cmidrule(lr){2-4}
      &Env-1&Env-2&Env-3\cr  
      \midrule  
      Straight &  0 & 0 & 0\\
      Random   &  0.003 & 0.001 & 0\\
      Our approach & 0.923 & 0.968 & 0.938 \\
      \bottomrule  
      \end{tabular}  
    \end{threeparttable}  
\end{table}  

\section{Discussion}
In this paper, a deep reinforcement learning based
framework is proposed for quadrotor autonomous obstacle avoidance.
Our framework has some characteristics as follows:

\begin{itemize}
    \item An unsupervised learning-based method for
    depth estimation is used for the environment perception
    module in our framework.
    It is novel to apply this method 
    to the quadrotor autonomous obstacle.
    In this paper, we train and test the module with raw
    data obtained by the quadrotor's on-board 
    monocular camera in the simulation.
    The training and testing results show that
    the model can effectively estimate 
    the distance of obstacles on the route of the quadrotor.
    However, due to the limitation of the quadrotor's flight ability, 
    the on-board camera is difficult to obtain 
    enough data under certain circumstances, 
    which results in the decline of depth estimation ability.
    
    \item The quadrotor mentioned in this paper only relies on 
    a monocular camera to obtain environment information, 
    which limits its observation ability and 
    makes it difficult to make effective 
    obstacle avoidance decisions.
    To solve this problem,
    we propose the D3RQN to learn the policy efficiently 
    with limited observations.
    It can learn the obstacle avoidance policy
    from previous observations rather than only from the current one.
    Compared with some other typical RL based methods, 
    our method has better learning efficiency 
    and test performance.

    \item Since application scenario transformation 
    is pervasive in UAV applications, 
    we tested the performance of the model after transformation.
    The test results show that our method can effectively 
    make proper obstacle avoidance decisions in the new scenarios
    after retraining the depth estimation network only,
    even though the obstacles in new scenarios are 
    different in appearance, shape, size and location arrangement.
    Besides, retraining the depth estimation network 
    in our framework only requires raw image sequences 
    without labels or groundtruth, 
    which is convenient to prepare. 
\end{itemize}

\section{Conclusion}

In this paper, the D3RQN framework is presented.
It can guide the quadrotor to achieve
autonomous obstacle avoidance only 
by inputting the image captured by an on-board monocular camera.
The training and testing results demonstrate
that the D3RQN has a better learning efficiency
and testing performance than some other approaches
such as double DQN, D3QN and double DRQN.
The test in different scenarios
shows that our framework has
good scenario migration ability.

In the future, the framework is going to have a more complex 
network structure to control the quadrotor with more
complex action space.
The improvement of training efficiency is also in 
consideration so that the framework can fit 
the limited on-board computing resource.






\bibliographystyle{elsarticle-num} 
\bibliography{AOA-DRQN_1}





\end{document}